\documentclass[11pt]{article}

\usepackage[a4paper,margin=3cm]{geometry}

\usepackage{cite}

\usepackage{graphicx,color}

\begin{document}

\title{\Large\bf Lighting up the Christmas tree: \\ high-intensity laser interactions with a nano-structured target}

\author{\normalsize A. Gonoskov, C. Harvey, A. Ilderton, F. Mackenroth, and M. Marklund \\
\it \normalsize Department of Applied Physics, Chalmers University of Technology, \\ \it \normalsize SE-412 96 Gothenburg, Sweden}

\maketitle

\begin{abstract}
	We perform a numerical study of the interaction of a high-intensity laser pulse with a nano-structured target. In particular, we study a target where the nano-structuring increases the absorption rate as compared to the flat target case. The transport of electrons within the target, and in particular in the nano-structure, is analysed. It is shown that it is indeed possible, using a terawatt class laser, to light up a nano-scale Christmas tree. Due to the form of the tree we achieve very strong edge fields, in particular at the top where the star is located. Such edge fields, as here located at ion rich spots, makes strong acceleration gradients possible. It also results in a nice, warm glow suitable for the holiday season. 
\end{abstract}

\section*{\normalsize Introduction}
Within the field of laser-plasma acceleration of ions~\cite{macchi},  the concept of nano-structured targets has caught the interest of the research community (see, e.g., \cite{nano} for a silicon target). The purpose of such structure is manifold, such as increasing the laser absorption rate of the target and increasing the strength of the accelerating fields \cite{macchi}. Motivated by this, and the festive season, we here analyse the properties of a particular nano-structured target, namely the Christmas tree (for an extensive review of Christmas trees, see e.g.\ Ref.\ \cite{tree}). Far from being a turkey, the Christmas tree and its ornaments have several properties making them suitable as a target design. It consists of several substructures allowing for higher absorption than flat targets~\cite{1,2,3,5}, it has several geometric features giving rise to complex edge fields, and it is generally considered to look great.


\section*{\normalsize Methods}
\begin{figure}
\begin{center}
\includegraphics[width=.4\textwidth]{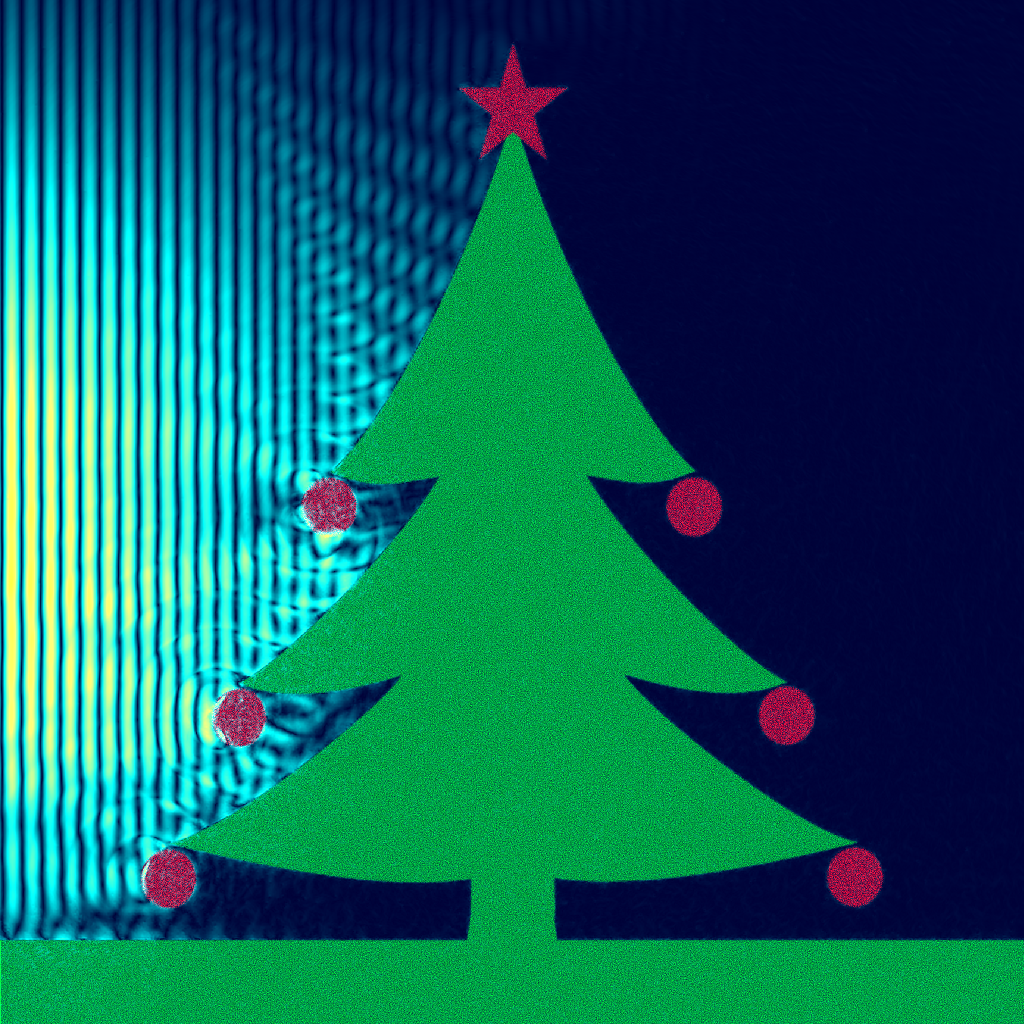}
\includegraphics[width=.4\textwidth]{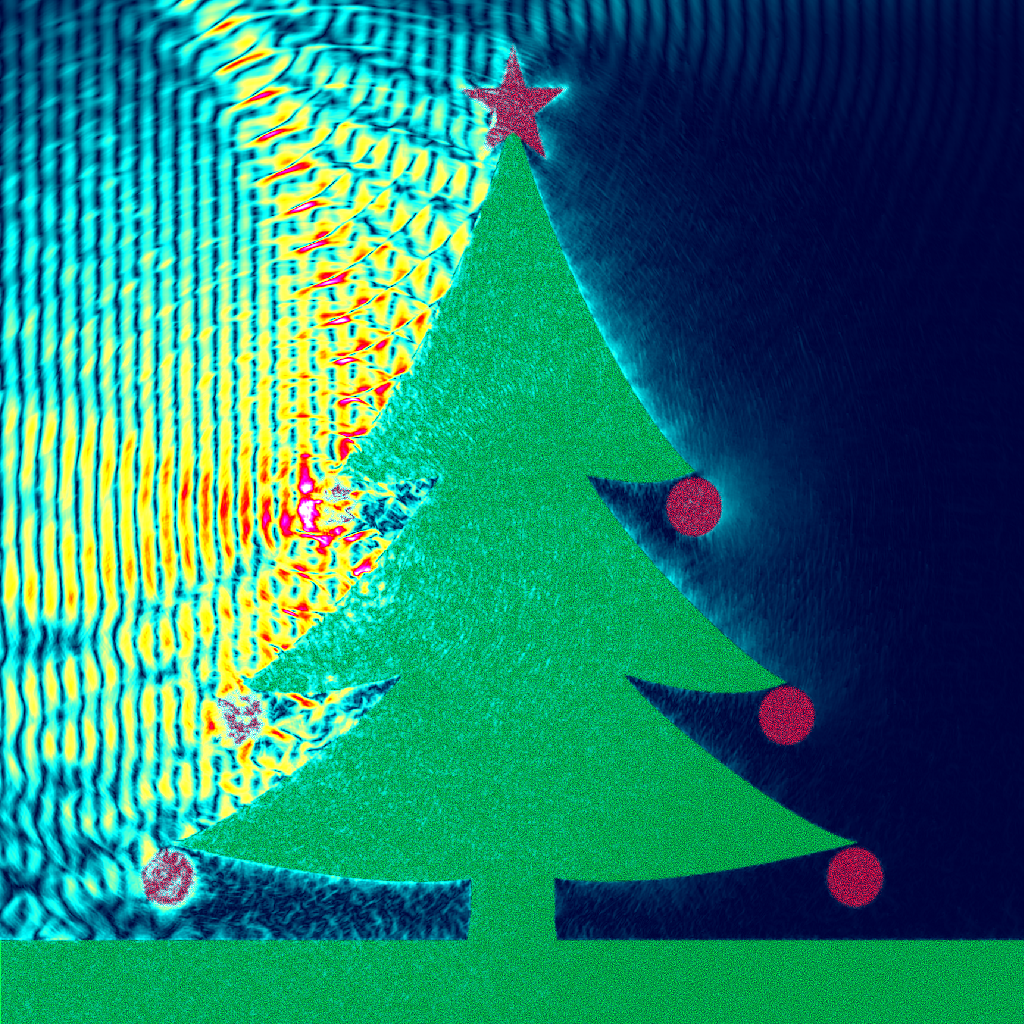} \\[1mm] 
\includegraphics[width=.4\textwidth]{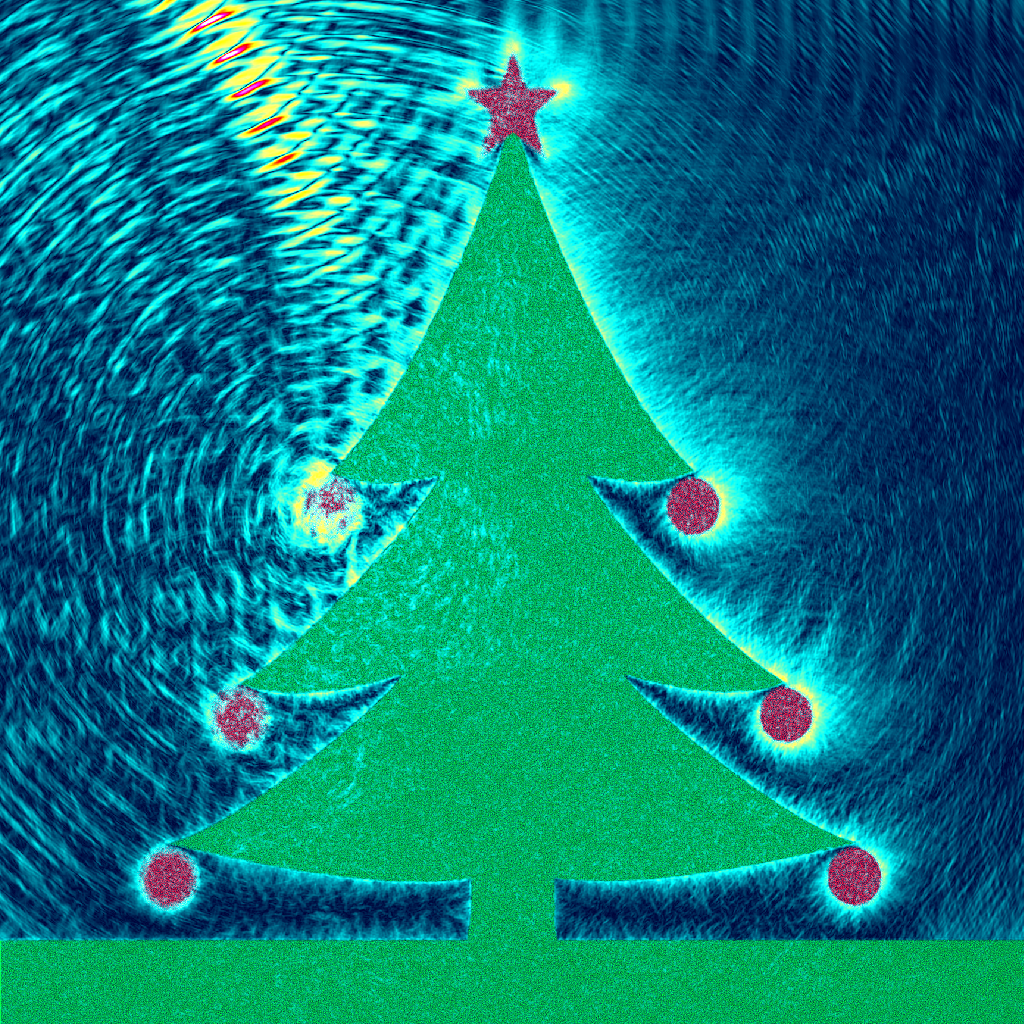}
\includegraphics[width=.4\textwidth]{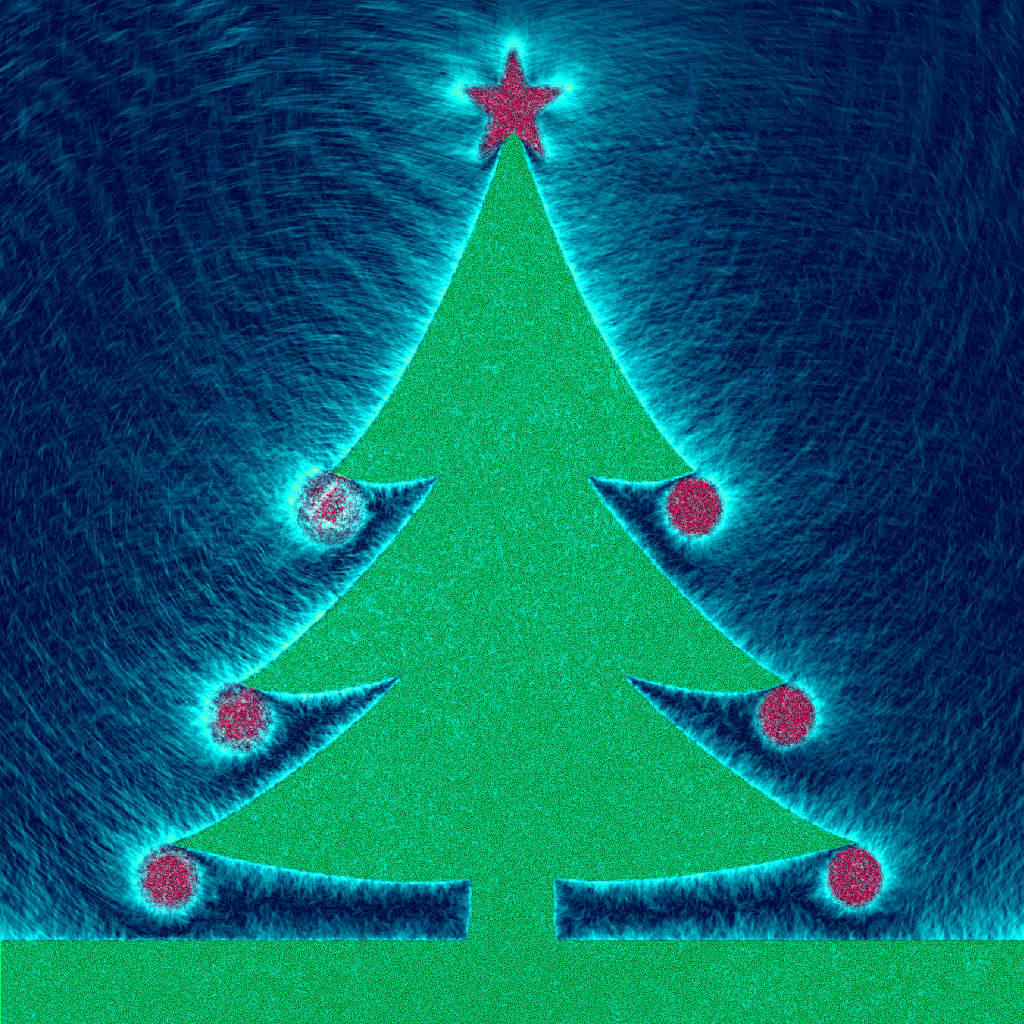}
\end{center}
\caption{\label{trees} A sequence from the simulation. In the upper left panel we see the initial pulse entering from the left. The upper right panel shows the target when the peak of the pulse enters. The lower left panel shows the outgoing, reflected pulse structure, while the lower right panel shows the lit up fir tree with its strong edge fields after the pulse has passed. }
\end{figure}

Our setup is as follows. We let a laser pulse impinge on a target with a Christmas tree, but flat otherwise, see the first panel in Fig.~\ref{trees}. Our target's main body consists of gold, while the ornaments (see also~\cite{4}) are made from a silica rich material, such as glass. This ensures that the ornaments, located where strong edge fields are created, could in principle act as sources for light ions -- here we focus our attention on the edge fields. The typical size of the ornaments is $\sim 1\, \mathrm{\mu m}$. We use a 2D particle-in-cell scheme to simulate the interaction of this target with a 35~fs laser pulse of 800~nm wavelength, focused to a 4~$\mu$m$\times$4~$\mu$m focal spot (FWHM for intensity, Gaussian profile), and having a total energy of 0.1~J.  The target density is, in critical units, 10 and 5 for the tree and the ornaments respectively. (We assume incomplete ionisation.) The time step in the simulation is $5.3 \times 10^{-17}$~s and the spatial region simulated, 50~$\mu$m$\times$50~$\mu$m, is represented by 2048$^2$ cells.

\section*{\normalsize Results and Discussion}
The results of the simulation are presented in Fig.~\ref{trees}. As is customary, we find that the Christmas tree conceals A~GIFT (Absorption Gain In Fir-tree Targets). The Christmas tree target allows for higher absorption rates than for flat TNSA setups. Further, it also presents a suitable tribute to the holiday season.
 
 The absorption gain is however not the main feature that makes these nano-structures of great interest. Instead, consider the edge fields located at the ion dots (ornaments) in red. After the laser pulse has passed, we clearly see that the christmas tree has been lit -- edge fields provide a warm fireside glow. We find particularly strong edge fields around the star, at the top of the Christmas tree. The presence of these edge fields implies that this seasonal setup may constitute a viable means of accelerating particles.

\section*{\normalsize Acknowledgements}
We are grateful for the spare time we had during which this research was conducted.

\end{document}